# Persistent room temperature photodarkening in Cu-doped β-Ga$_2$O$_3$

J. Jesenovec[1-2], C. Pansegrau,[3] M.D. McCluskey,[1-3,*] J.S. McCloy,[1-2] T.D. Gustafson,[4] L.E. Halliburton,[5] and J.B. Varley[6]

[1]*Institute of Materials Research, Washington State University, Pullman, Washington, USA 99164-2711*
[2]*Materials Science & Engineering Program, Washington State University, Pullman, Washington, 99164, USA*
[3]*Department of Physics and Astronomy, Washington State University, Pullman, Washington 99164-2814, USA*
[4]*Department of Engineering Physics, Air Force Institute of Technology, Wright-Patterson Air Force Base, Ohio 45433, USA*
[5]*Department of Physics and Astronomy, West Virginia University, Morgantown, West Virginia, 26506-6315, USA*
[6]*Lawrence Livermore National Laboratory, Livermore, California 94551-0808, USA*

*Corresponding author:* mattmcc@wsu.edu

ORCID:
Jesenovec: 0000-0002-5937-6657
Pansegrau: 0000-0003-4800-7474
McCluskey: 0000-0002-0786-4106
McCloy: 0000-0001-7476-7771
Gustafson: 0000-0002-5519-4610
Halliburton: 0000-0002-8228-5024
Varley: 0000-0002-5384-5248

**Abstract:** β-Ga$_2$O$_3$ is an ultra-wide bandgap semiconductor with emerging applications in power electronics. The introduction of acceptor dopants yields semi-insulating substrates necessary for thin-film devices. In the present work, exposure of Cu-doped β-Ga$_2$O$_3$ to UV light > 4 eV is shown to cause large, persistent photo-induced darkening at room temperature. Electron paramagnetic resonance spectroscopy indicates that light exposure converts Cu$^{2+}$ to Cu$^{3+}$, a rare oxidation state that is responsible for the optical absorption. The photodarkening is accompanied by the appearance of O-H vibrational modes in the infrared spectrum. Hybrid function calculations show that Cu acceptors can favorably complex with hydrogen donors incorporated as interstitial (H$_i$) or substitutional (H$_O$) defects. When Cu$_{Ga}$-H$_O$ complexes absorb light, hydrogen is released, contributing to the observed Cu$^{3+}$ species and O-H modes.

Monoclinic β-Ga$_2$O$_3$ has an ultra-wide bandgap of 4.8 eV and predicted breakdown field of 8 MV/cm, making it a potentially important material for high voltage, high power devices.[1–7] Insulating β-Ga$_2$O$_3$ substrates, required for devices such as field effect transistors, can be achieved by doping with acceptors such as Fe, Mg, and Zn.[8–12] Attempts at obtaining single crystal β-Ga$_2$O$_3$ with another possible acceptor, Cu, has been reported for Czochralski (CZ) growth at 0.2 mol.% Cu, but these trials yielded no successful incorporation due to high Cu vapor pressure.[8] Hydrogen is an omnipresent contaminant in bulk β-Ga$_2$O$_3$ crystals that can compensate or passivate acceptor dopants,[13,14] gallium vacancies,[15] and divacancies.[16,17]

In this Letter, we report persistent room temperature photodarkening in Cu-doped β-Ga$_2$O$_3$. Such a color change caused by light exposure, or photochromism, is rare in semiconductors. It has been observed in SrTiO$_3$ at low temperatures and attributed to changes in the charge state of transition-metal impurities.[18] In that work, the effect was not persistent at room temperature because electrons have sufficient thermal energy to surmount kinetic barriers and return to the ground state. More recently, persistent photoconductivity (PPC) was observed in annealed SrTiO$_3$ at room temperature.[19–21] The PPC effect is accompanied by optical absorption from Fe impurities.[22,23]

Cu-doped β-Ga$_2$O$_3$ single crystals (nominally doped at 0.25 at.% Cu on the metal site) were grown from the melt by similar methods to those previously published.[24–26] High purity (5N=99.999%) Ga$_2$O$_3$ powder (ABSCO Limited, Haverhill, Suffolk, UK) was used as source material. Cu$_2$O powder (2N purity, Alfa Aesar) was added to achieve cation doping. Crystals were grown in a 70 mm height × 70 mm outer diameter iridium crucible heated by a 25 kHz radio frequency inductive heating coil. A mixed Ar+O$_2$ gas was used during the melting and growth with varying O$_2$ partial pressure.[27] During the growth of β-Ga$_2$O$_3$:Cu crystal, the gas overpressure was 20–34 kPa and the O$_2$ concentration was 10–11%. Crystals were grown both by the CZ method with a 2 – 5 mm/hr pull rate and 2 rpm rotation of the crucible, and by vertical gradient freeze (VGF) with no rotation while cooling at 1 – 2°C min$^{-1}$. (100)-orientation samples were obtained from the VGF and CZ portions of the growth for the optical and magnetic resonance studies respectively. Glow discharge mass spectrometry (GDMS), performed at EAG Laboratory (California, USA), indicated a Cu density of ~10$^{19}$ cm$^{-3}$, along with Fe and Si impurities each at ~10$^{18}$ cm$^{-3}$. Details of the growth, secondary phases, and luminescence of the VGF samples are given in Ref. 26.

Optical transmission measurements in the ultraviolet (UV) through near infrared were performed with a Cary 5 UV-Vis-NIR spectrometer at room temperature for wavelengths 200 – 3000 nm on cleaved and as-grown samples of β-Ga$_2$O$_3$:Cu with thicknesses 0.5 – 1 mm. Low-temperature (10.6 K, achieved using a Janis closed-cycle cryostat) infrared (IR) absorption spectra with 0.5 cm$^{-1}$ resolution were obtained with a Bomem DA8 vacuum FTIR spectrometer with an InSb detector and KBr beamsplitter. A Bruker EMX spectrometer was operated near 9.38 GHz to collect electron paramagnetic resonance (EPR) spectra. Data were collected at room temperature and at 40 K using an Oxford Instruments ESR-900 cryostat with helium as the cooling gas. Samples were photodarkened by exposing to a 275 nm light emitting diode (LED, Inolux C39CTKU1) operating at ~5.45 V and ~0.1 A.

To study the decay of photodarkening, samples were placed in an open silica tube exposed to air and annealed in a tube furnace at 400°C for various durations. Samples were quenched immediately in air after annealing. These samples were studied against a reference, which was photodarkened and then left at room temperature for several days in ambient light. Resistivity was measured as a function of time after photodarkening, utilizing a hotplate at 400°C in air for erasing.

For resistivity measurements, ohmic contacts were placed in a two-point configuration using 50-50 wt.% Ga-In. These samples were then annealed at 950°C for 15 min and a small amount of Ga-In was placed on top of the contact again.[24]

Cu-related defect formation energies ($E^f$), thermodynamic and vertical transition levels, and migration barriers were calculated using the Heyd-Scuseria-Ernzhof screened hybrid functional (HSE06) and projector-augmented wave (PAW) approach as implemented in the VASP code.[28–31] All calculations were performed using supercells with 160-atoms (a 3×4×1 repetition of the 20-atom unit cell) with the same computational approach and finite size corrections for the formation energies and thermal and vertical transition energies and as detailed in previous publications.[32–35] We additionally account for the effects of limiting phases in the calculated chemical potential of Cu dopants as a function of conditions, finding CuO (calculated $\Delta H$[CuO] = –1.64 eV / formula unit) and $Ga_2Cu$ (calculated $\Delta H$[$Ga_2Cu$] = –0.31 eV / formula unit) to be the solubility-limiting phases in the O-rich and Ga-rich limits, respectively. Migration barriers were calculated using the climbing image nudged elastic band approach[36] using 5 images and a force tolerance of 0.03 eV/Å.

Figure 1(a) shows a UV/visible transmission spectrum of a sample exposed to 275 nm light. Upon light exposure, strong absorption bands at 379 nm and 476 nm (3.27 eV and 2.61 eV) are observed. The absorption resulted in a dark red appearance of the sample. Photodarkening was persistent at room temperature, decaying by roughly half over 14 days. The electrical resistance was reduced by ~50% when darkened. Annealing in air at 400°C for 5 min [Fig. 1(b)] returns the sample to its as-grown color and the resistance recovered its initial value of $10^{11}$ Ω (estimated resistivity $10^{10}$ Ω·cm). Resistance recovery occurred within 30 s at 400°C with minimal further recovery after annealing for another 5 min.[26]

Concomitant with photodarkening, O-H peaks appeared in the IR spectrum. The as-grown sample only showed a small ZnH peak (3487.6 cm$^{-1}$) due to residual Zn and H impurities, with GDMS indicating ~10$^{16}$ Zn atoms/cm$^3$.[14] After exposure to 275 nm light, a series of IR absorption peaks corresponding to O-H bond stretching vibrational modes is observed. Two IR peaks appear at 3416.1 and 3438.9 cm$^{-1}$ [Fig. 2(a)] each with a full width at half maximum (FWHM) of 2.5 cm$^{-1}$. Two sharper peaks, with FWHMs near the instrumental resolution of 1 cm$^{-1}$, are observed at 3466.6 and 3484.6 cm$^{-1}$. All four peaks disappear after annealing at 400°C. Following the anneal, photodarkening a second time restores the IR peaks to their original strengths. A calculated frequency of 3465 cm$^{-1}$ for the favorable Cu$_{\text{GaII}}$-H configuration, analogous to those observed in Mg-doped samples[13] with the same anharmonic and experimental corrections, suggests these peaks are due to CuH complexes.

The IR spectra also contain a peak at 5147.6 cm$^{-1}$, which is due to an electronic transition of isolated Ir$^{4+}$.[13,37] The appearance of hydrogen peaks is accompanied by the decrease in the Ir$^{4+}$ peak [Fig. 2(b)]. This reduction in peak intensity is due to Ir$^{4+}$ capturing an electron, turning it into IR-inactive Ir$^{3+}$. Annealing the sample at 400°C restores the Ir$^{4+}$ to its original state.

EPR spectra taken with the magnetic field along $a$ for Cu$^{2+}$ and Cu$^{3+}$ are shown in Fig. 3. Cu$^{2+}$ (3d$^9$) is the dominant spectrum observed in an as-grown crystal at 40 K and is characterized by a four-line hyperfine pattern caused by interaction with the copper nuclei. Following exposure to a 275 nm LED, the Cu$^{2+}$ signal is reduced by about a factor of four, and the Cu$^{3+}$ spectrum appears. Cu$^{3+}$ (3d$^8$) is a spin $S = 1$ system that gives two lines split by a large zero-field splitting that are additionally split into four hyperfine lines by the copper nuclei. Heating the crystal to 400°C and holding for 2 min is sufficient to return the EPR to the as-grown crystal state, consistent with the IR and UV/visible results. A second, smaller $Cu^{2+}$ EPR spectrum is present in the as-grown crystal and observed at room temperature. It appears under the second $Cu^{2+}$ line in Fig. 3(a).

The $Cu^{3+}$ oxidation state is rare compared to the much more common $Cu^0$, $Cu^+$, and $Cu^{2+}$. $Cu^{3+}$ has been observed in $Al_2O_3$ and has similar EPR spectrum to what we observe, as well as absorption bands at 590 nm, 470 nm, and 330 nm due to internal electronic transitions from the $^3A_2$ ground state to excited states.[38] All the Cu dopants in the as-grown $Al_2O_3$:Cu crystals were in the $Cu^{3+}$ state, with no $Cu^{2+}$ observed. In the present case of $Ga_2O_3$, however, $Cu^{3+}$ is only observed after illumination. Its persistence in the $Cu^{3+}$ state is attributed to the photoexcited electron becoming trapped by $Ir^{4+}$ and possibly other defects.

The results discussed so far were obtained with 275 nm (4.51 eV) light, which is just below the bandgap. To determine the optical threshold for photodarkening, a set of samples was exposed to LEDs of different wavelengths. The photodarkening spectrum was calculated as

Absorbance = $\log_{10}(T_0/T)$

where $T_0$ is the transmission before exposure to light (the reference) and $T$ was the transmission after exposure. As shown in Fig. 4, the optical threshold is ~300 nm (4.1 eV), which is clearly below the bandgap.

Hybrid functional calculations indicate that Cu prefers to occupy the octahedral Ga(II) site and has a (0/-) acceptor level 2.3 eV above the VBM. In the negative charge state, denoted $Cu^{2+}$ or $Cu_{Ga}^-$, the energy for the Ga(II) site is 0.3 eV lower than the Ga(I) site, which exhibits a 1.6 eV acceptor level. These deep acceptor levels are consistent with the observation of $S=½$ $Cu^{2+}$ centers in the as-grown material (formation energies for Cu and other defects are shown in Fig. S1). The calculated absorption for exciting an electron from $Cu^{2+}$ to the conduction-band minimum (vertical transition in a configuration coordinate diagram) is 3.3 eV (376 nm) for $Cu_{GaII}$ and 3.8 eV (328 nm) for $Cu_{GaI}$. Both $Cu_{Ga}^0$ configurations result in a $S=1$ state ($Cu^{3+}$). These energy thresholds are lower than the ~4.1 eV onset for significant photodarkening (Fig. 4).

The "simple Cu model" also does not explain the appearance of hydrogen modes. However, the calculations indicate that a copper acceptor ($Cu^{2+}$) can pair with substitutional hydrogen donor ($H_O^+$) on the O(I) site to form a neutral complex, $(Cu_{GaII}\text{-}H_{OI})^0$. ($H_O^+$ can also be described as an $H^-$ ion trapped in a doubly ionized oxygen vacancy, $V_O^{++}$). In this complex, $H_{OI}$ would not give rise to an O-H mode and is therefore "hidden" from IR spectroscopy. $H_O$ is known to be a source of hidden hydrogen in a number of other oxides that may be released upon annealing[39] or photoexcitation.[20,21] We propose that $Cu_{GaII}\text{-}H_{OI}$ and isolated $Cu_{GaII}$ are responsible for the $Cu^{2+}$ EPR signatures in Fig. 3(a).

From the configuration coordinate diagram in Fig. 5(a), the expected absorption threshold for exciting the electron from $Cu_{Ga}\text{-}H_{OI}$ to the conduction band is 3.82 eV, consistent with experiment (Fig. 4). This photoionization process is given by

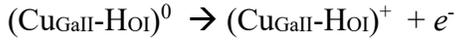
$(Cu_{GaII}\text{-}H_{OI})^0 \rightarrow (Cu_{GaII}\text{-}H_{OI})^+ + e^-$

As the ionized complex not stable [Fig. 5(b)], it becomes more energetically favorable for the proton to leave the substitutional site via the process

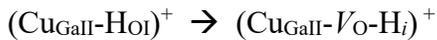
$(Cu_{GaII}\text{-}H_{OI})^+ \rightarrow (Cu_{GaII}\text{-}V_O\text{-}H_i)^+$

where the $H_i$ moves offsite to an adjacent O(I). In the photoionized $(Cu_{GaII}\text{-}V_O\text{-}H_i)^+$ defect, Cu is in the $Cu^{3+}$ state ($S = 1$) and therefore may contribute to the EPR spectrum in Fig. 3(b).

The $Cu_{Ga}\text{-}V_O\text{-}H_i$ complex can undergo further dissociation, resulting in $Cu_{Ga}\text{-}V_O$ and mobile $H_i$. The liberated, mobile $H_i$ can then form complexes with point defects present in the lattice. We propose that the IR peaks observed in Fig. 2(a) are due to $Cu_{Ga}\text{-}H_i$ complexes, but cannot exclude the possibility that some of the lines are due to complexes between hydrogen and other defects with lower concentrations than Cu.

In conclusion, optical and EPR experiments show that photons of energy > 4 eV cause a $Cu^{2+} \rightarrow Cu^{3+}$ transition accompanied by persistent photodarkening. The optical absorption is attributed to an internal electronic transition of the $Cu^{3+}$ ion. Hybrid functional calculations indicate that the Cu in question may be an isolated acceptor ($Cu_{GaII}$) or more likely part of an acceptor-donor pair involving hydrogen and/or oxygen vacancies. The $Cu_{GaII}$-$H_{OI}$ defect model, in particular, explains the appearance of O-H vibrational modes after illumination and has a calculated absorption threshold that agrees with experiment. Further experimental characterization will be required to confirm the defect processes underpinning this interesting effect.

This research was supported by Air Force Office of Scientific Research under award number FA9550-18-1-0507 monitored by Dr. Ali Sayir. Any opinions, finding, and conclusions or recommendations expressed in this material are those of the authors and do not necessarily reflect the views of the United States Air Force. M.D.M. and C.P. acknowledge support by the U.S. Department of Energy, Office of Basic Energy Sciences, Division of Materials Science and Engineering under Award No. DE-FG02-07ER46386. T.D.G. was supported at the Air Force Institute of Technology by an NRC Research Associateship Award. This work was partially performed (J.B.V.) under the auspices of the U.S. DOE by Lawrence Livermore National Laboratory under contract DE-AC52-07NA27344, and supported by the Critical Materials Institute, an Energy Innovation Hub funded by the U.S. DOE, Office of Energy Efficiency and Renewable Energy, Advanced Manufacturing Office.

# REFERENCES

[1] M. Baldini, Z. Galazka, and G. Wagner, Mater. Sci. Semicond. Process. **78**, 132 (2018).

[2] M. Bartic, C.-I. Baban, H. Suzuki, M. Ogita, and M. Isai, J. Am. Ceram. Soc. **90**, 2879 (2007).

[3] E. Farzana, Z. Zhang, P.K. Paul, A.R. Arehart, and S.A. Ringel, Appl. Phys. Lett. **110**, 202102 (2017).

[4] A. Bhattacharyya, P. Ranga, M. Saleh, S. Roy, M. A. Scarpulla, K. G. Lynn, and S. Krishnamoorthy, IEEE J. Electron Devices Soc. **8**, 286 (2020).

[5] Y. Li, T. Tokizono, M. Liao, M. Zhong, Y. Koide, I. Yamada, and J.-J. Delaunay, Adv. Funct. Mater. **20**, 3972 (2010).

[6] M. Higashiwaki, K. Konishi, K. Sasaki, K. Goto, K. Nomura, Q.T. Thieu, R. Togashi, H. Murakami, Y. Kumagai, B. Monemar, A. Koukitu, A. Kuramata, and S. Yamakoshi, Appl. Phys. Lett. **108**, 133503 (2016).

[7] M. Higashiwaki and G.H. Jessen, Appl. Phys. Lett. **112**, 060401 (2018).

[8] Z. Galazka, K. Irmscher, R. Schewski, I.M. Hanke, M. Pietsch, S. Ganschow, D. Klimm, A. Dittmar, A. Fiedler, T. Schroeder, and M. Bickermann, J. Cryst. Growth **529**, 125297 (2020).

[9] A.Y. Polyakov, N.B. Smirnov, I.V. Shchemerov, S.J. Pearton, F. Ren, A.V. Chernykh, and A.I. Kochkova, Appl. Phys. Lett. **113**, 142102 (2018).

[10] J. Jesenovec, S.E. Karcher, J.B. Varley, and J.S. McCloy, Press J. Appl. Phys. (2021).

[11] M.D. McCluskey, J. Appl. Phys. **127**, 101101 (2020).

[12] T.D. Gustafson, J. Jesenovec, C.A. Lenyk, N.C. Giles, J.S. McCloy, M.D. McCluskey, and L.E. Halliburton, J. Appl. Phys. **129**, 155701 (2021).

[13] J.R. Ritter, J. Huso, P.T. Dickens, J.B. Varley, K.G. Lynn, and M.D. McCluskey, Appl. Phys. Lett. **113**, 052101 (2018).

[14] C. Pansegrau, J. Jesenovec, J.S. McCloy, and M.D. McCluskey, Appl. Phys. Lett. **119**, 102104 (2021).

[15] P. Weiser, M. Stavola, W.B. Fowler, Y. Qin, and S. Pearton, Appl. Phys. Lett. **112**, 232104 (2018).

[16] C. Zimmermann, E.F. Verhoeven, Y.K. Frodason, P.M. Weiser, J.B. Varley, and L. Vines, J. Phys. Appl. Phys. **53**, 464001 (2020).

[17] Y.K. Frodason, C. Zimmermann, E.F. Verhoeven, P.M. Weiser, L. Vines, and J.B. Varley, Phys. Rev. Mater. **5**, 025402 (2021).

[18] B.W. Faughnan, Phys. Rev. B **4**, 3623 (1971).

[19] M.C. Tarun, F.A. Selim, and M.D. McCluskey, Phys. Rev. Lett. **111**, 187403 (2013).

[20] V.M. Poole, J. Huso, and M.D. McCluskey, J. Appl. Phys. **123**, 161545 (2018).

[21] Z. Zhang and A. Janotti, Phys. Rev. Lett. **125**, 126404 (2020).

[22] V.M. Poole, M.D. McCluskey, J. Dashdorj, and M.E. Zvanut, MRS Online Proc. Libr. **1792**, 706 (2015).

[23] A. Viernstein, M. Kubicek, M. Morgenbesser, G. Walch, G.C. Brunauer, and J. Fleig, Adv. Funct. Mater. **29**, 1900196 (2019).

[24] M. Saleh, A. Bhattacharyya, J.B. Varley, S. Swain, J. Jesenovec, S. Krishnamoorthy, and K. Lynn, Appl. Phys. Express **12**, 085502 (2019).

[25] M. Saleh, J.B. Varley, J. Jesenovec, A. Bhattacharyya, S. Krishnamoorthy, S. Swain, and K. Lynn, Semicond. Sci. Technol. **35**, 04LT01 (2020).

[26] J. Jesenovec, C. Remple, J. Huso, B. Dutton, P. Toews, M.D. McCluskey, and J.S. McCloy, J. Cryst. Growth **578**, 126419 (2022).

[27] Z. Galazka, R. Uecker, D. Klimm, K. Irmscher, M. Naumann, M. Pietsch, A. Kwasniewski, R. Bertram, S. Ganschow, and M. Bickermann, ECS J. Solid State Sci. Technol. **6**, Q3007 (2016).

[28] J. Heyd, G.E. Scuseria, and M. Ernzerhof, J. Chem. Phys. **118**, 8207 (2003).

[29] P.E. Blöchl, Phys. Rev. B **50**, 17953 (1994).

[30] G. Kresse and J. Furthmüller, Comput. Mater. Sci. **6**, 15 (1996).

[31] G. Kresse and J. Furthmüller, Phys. Rev. B **54**, 11169 (1996).

[32] C. Freysoldt, B. Grabowski, T. Hickel, J. Neugebauer, G. Kresse, A. Janotti, and C.G. Van de Walle, Rev. Mod. Phys. **86**, 253 (2014).

[33] M.E. Ingebrigtsen, A.Yu. Kuznetsov, B.G. Svensson, G. Alfieri, A. Mihaila, U. Badstübner, A. Perron, L. Vines, and J.B. Varley, APL Mater. **7**, 022510 (2019).

[34] Y.K. Frodason, K.M. Johansen, L. Vines, and J.B. Varley, J. Appl. Phys. **127**, 075701 (2020).

[35] T. Gake, Y. Kumagai, C. Freysoldt, and F. Oba, Phys. Rev. B **101**, 020102 (2020).

[36] G. Henkelman, B.P. Uberuaga, and H. Jónsson, J. Chem. Phys. **113**, 9901 (2000).

[37] C.A. Lenyk, N.C. Giles, E.M. Scherrer, B.E. Kananen, L.E. Halliburton, K.T. Stevens, G.K. Foundos, J.D. Blevins, D.L. Dorsey, and S. Mou, J. Appl. Phys. **125**, 045703 (2019).

[38] W.E. Blumberg, J. Eisinger, and S. Geschwind, Phys. Rev. **130**, 900 (1963).

[39] Y. Qin, M. Stavola, W.B. Fowler, P. Weiser, and S.J. Pearton, ECS J. Solid State Sci. Technol. **8**, Q3103 (2019).

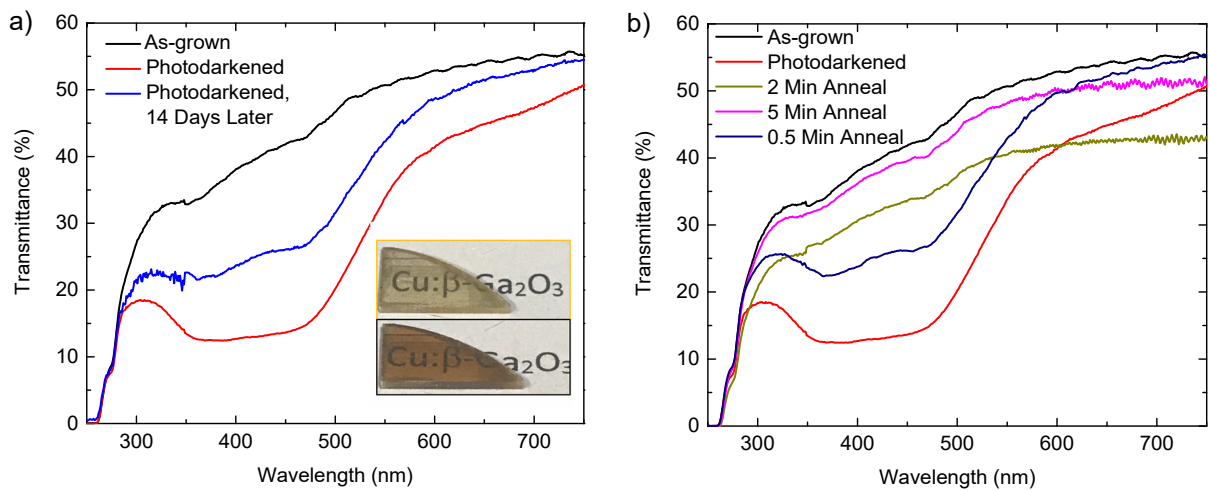

Figure 1. (a) Transmission spectra of as-grown and photodarkened β-Ga$_2$O$_3$, then remeasured 14 days later. Inset shows a sample as-grown and after photodarkening. (b) As-grown sample photodarkened and then annealed at 400°C, showing an almost complete return to as-grown transmission after a 5 minute anneal. Note that the spectrometer detector showed some drift that affected the value of the transmittance, especially for the 2 min anneal.

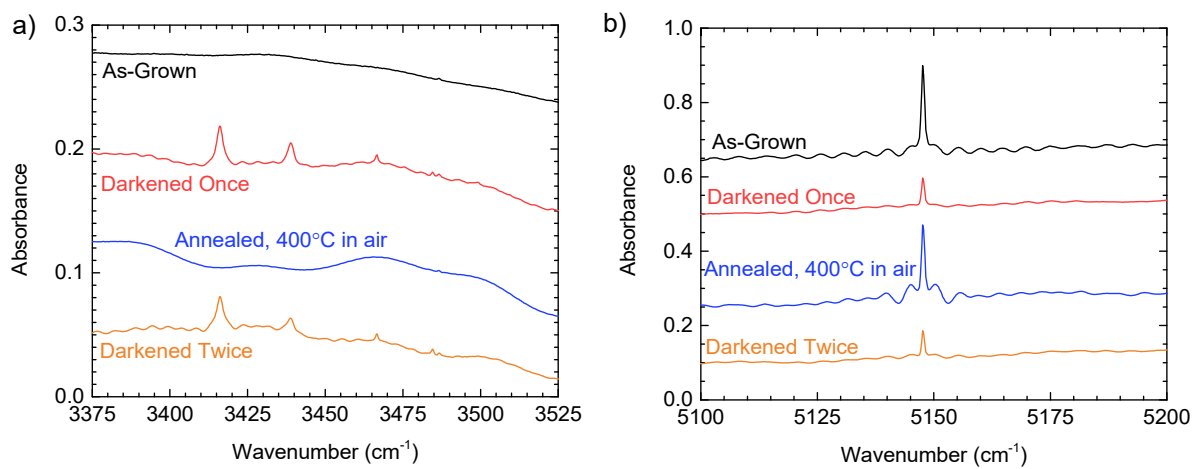

Figure 2. Low-temperature (11 K) IR absorption spectra of β-$Ga_2O_3$:Cu (thickness 0.49 mm). (a) O-H absorption peaks, attributed to CuH complexes, appeared after photodarkening with 275 nm LED. Annealing in air eliminated the peaks as well as the photodarkening. Photodarkening a second time brought the peaks back. (b) Corresponding changes in the $Ir^{4+}$ absorption peak. Spectra are offset vertically for clarity. Sample thickness = 0.49 mm.

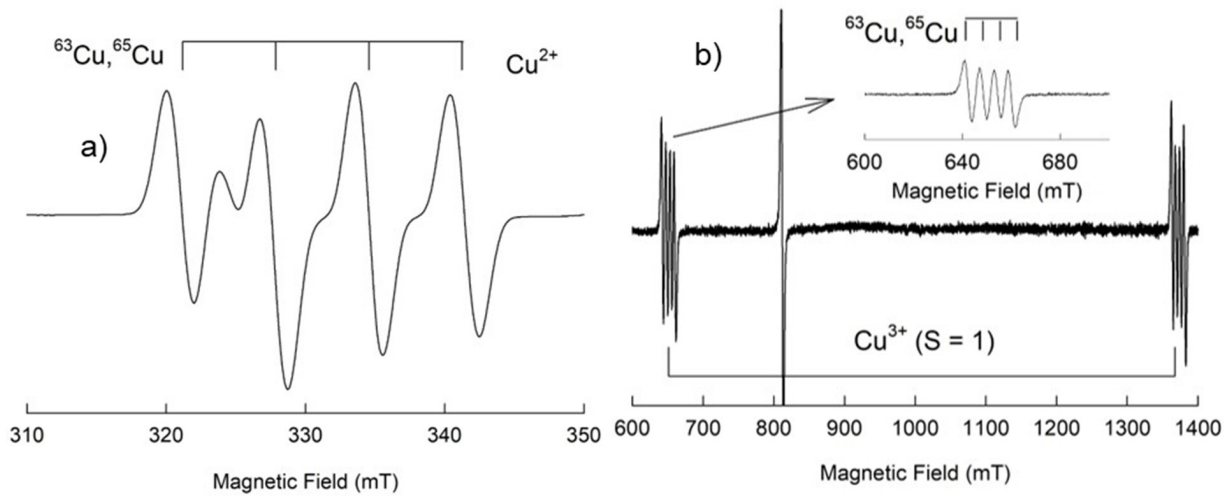

Figure 3. EPR spectra for (a) $Cu^{2+}$ at 40 K and (b) $Cu^{3+}$ at room temperature. Both were taken with the magnetic field along the *a* direction. The $Cu^{3+}$ spectrum appears following UV exposure and is eliminated by heating the crystal to 400 °C. (The line near 800 mT is due to $Fe^{3+}$).

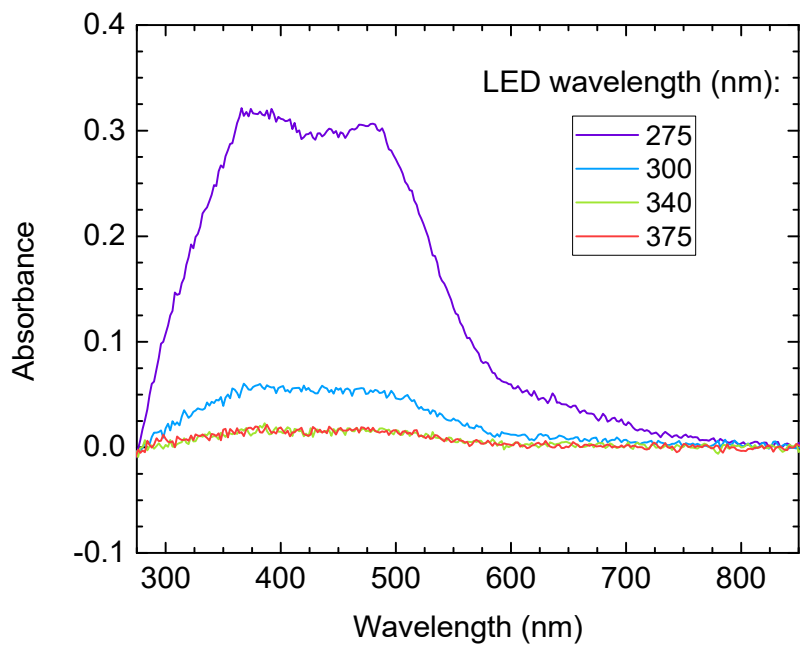

Figure 4. Absorbance spectra of β-Ga$_2$O$_3$:Cu samples that were exposed to LEDs of different wavelengths. Significant photodarkening is observed for the 300 and 275 nm LEDs. Sample thicknesses were 0.67 mm (same sample for 375 and 275 nm exposures), 0.49 mm (340 nm exposure, same sample as Fig. 2), and 0.45 mm (300 nm exposure).

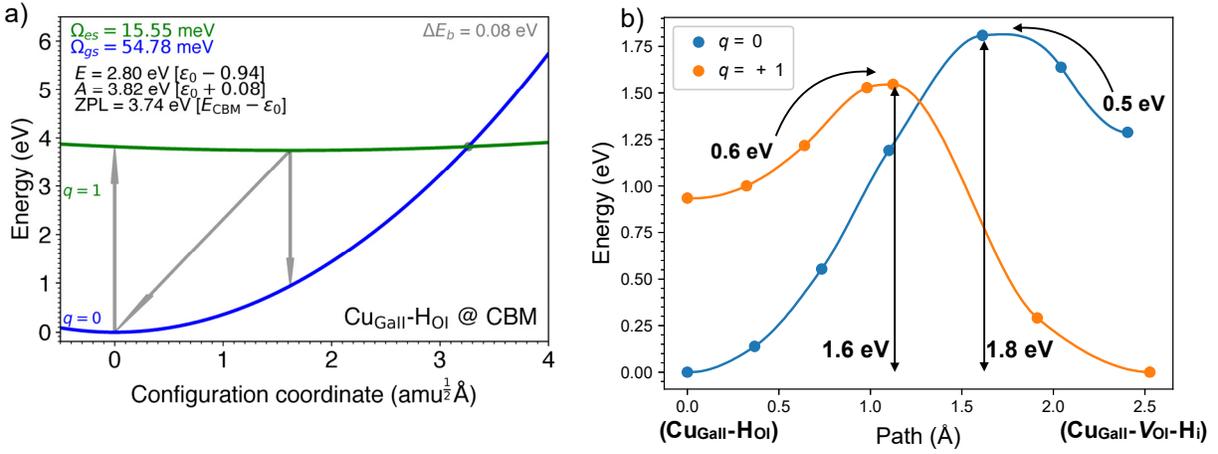

Figure 5. (a) Configuration-coordinate diagram for $Cu_{GaII}$-$H_{OI}$. The curves are for neutral ($q = 0$) and positive ($q = 1$) charge states of the defect relative to the host. Energies are indicated for excited-state and ground-state relaxation ($\Omega_{es}$ and $\Omega_{gs}$), absorption ($A$, vertical up arrow), emission ($E$, vertical down arrow), and zero-phonon line (ZPL, diagonal down arrow). (b) Reaction path for ($Cu_{GaII}$-$H_{OI}$) → ($Cu_{GaII}$-$V_{OI}$-$H_i$), shown for the neutral and positive charge states.